\documentclass[a4paper,11pt]{article}
\pdfoutput=1 

\usepackage{jheppub} 

\usepackage[T1]{fontenc} 

\usepackage{mathtools}

\title{\boldmath Quantum Tunneling from the Charged Non-Rotating BTZ Black Hole with GUP}

\author[a]{Jafar Sadeghi}
\author[b]{Vahid Reza Shajiee}

\affiliation[a]{Department of Physics, University of Mazandaran,\\Babolsar, Iran}
\affiliation[b]{Young Researchers and Elite Club, Mashhad Branch, Islamic Azad University,\\Mashhad, Iran}

\emailAdd{pouriya@ipm.ir}
\emailAdd{v.shajiee@mshdiau.ac.ir}


\abstract{In the present paper, the quantum corrections to the temperature, entropy and specific heat capacity of the charged non-rotating BTZ black hole are studied by generalized uncertainty principle in tunneling formalism. It is shown that quantum corrected entropy would be of the form of predicted entropy in quantum gravity theories like string theory and loop quantum gravity.}

\begin{document}
\maketitle
\flushbottom

\section{Introduction}
\label{sec:1}

In the decade of the seventies, Bekenstein presented some analogies between black hole physics and thermodynamics~\cite{Bekenstein:1972tm,Bekenstein:1973ur,Bekenstein:1974ax}, and Hawking with collaborators introduced the four laws of black hole mechanics~\cite{Bardeen:1973gs}. In the same decade, Hawking showed that black holes radiate thermally through quantum vacuum fluctuations near their event horizons~\cite{Hawking:1974rv,Hawking:1974sw}. After such works, many efforts in theoretical physics had been accomplished to investigate thermodynamic properties of black holes. Despite many efforts, there exist numerous unsolved problems in the black holes physics; namely the microscopic origin of black holes entropy is not fully understood.

In the last half century, the quantum gravity problem has been one of the most important issues for theoretical physicists. In this route, existence of a minimal length and also a quantum corrected entropy,
\begin{equation}\label{1.1}
 \hat{S} = S + \alpha \ln S + \beta S^{-1} + \gamma ,
\end{equation}
have been proposed by quantum gravity candidate theories like string theory and loop quantum gravity~\cite{Carlip:2000nv,Kaul:2000kf,Medved:2004yu,Myung:2003ra,Meissner:2004ju}. In the equation \eqref{1.1}, the coefficients $\alpha$, $\beta$ and $\gamma$ are parameters related to the theory.

In view of the foregoing, it is natural to ask how to examine a quantum gravity candidate theory. A simple answer would be that an eligible theory should have good descriptions, with some likely corrections, of physical objects such as black holes. Therefore, one may namely investigate the thermodynamic properties of black holes in such a theory. One approach to this end is the generalized uncertainty principle (GUP)\footnote{For recent works on GUP and related topics, see~\cite{Faizal:2014rwa,Pramanik:2014mma,Ali:2015ola,Garattini:2015aca,Faizal:2016zlo,Faizal:2014pia,Faizal:2014dua,Faizal:2014rha,Faizal:2014mfa,Faizal:2014mba}.} given by~\cite{Ali:2009zq}
\begin{equation}\label{1.2}
  \Delta x \Delta p \geq \hbar (1-\frac{\lambda l_{p}}{\hbar} \Delta p + \frac{\lambda^{2} l_{p}^{2}}{\hbar^{2}} \Delta p^{2}) ,
\end{equation}
where $\lambda$ is a dimensionless positive parameter, $l_{p}=\sqrt{\frac{\hbar G}{c^{3}}}=\frac{M_{p} G}{c^{2}} \approx 10^{-35} (m)$, $M_{p}=\sqrt{\frac{\hbar c}{G}}$ and $c$ are Planck length, Planck mass and velocity of light, respectively. In obtaining the equation \eqref{1.2} from that of~\cite{Ali:2009zq}, it is assumed that $<p> \sim 0$ and consequently $<p^{2}> \sim \Delta p^{2}$. From other point of view, due to the Planck length is proportional to Newton coupling constant, the quantum corrections in equation \eqref{1.2} may be considered as gravitational effects. Black hole physics and string theory suggest a GUP including a quadratic term in the momenta while doubly special relativity (DSR) suggests one that is linear in the momenta~\cite{Ali:2011fa}. So, it is appropriate to use the most general form of GUP with linear and quadratic terms, this is the physical motivation for using the specific form of GUP in the equation \eqref{1.2}. On the other hand, the Hawking radiation may be considered as quantum tunneling across the event horizon~\cite{Parikh:1999mf}. One approach to such an investigation is Hamilton-Jacobi method wherein it has been used from WKB approximation to determine the particles's tunneling rate~\cite{Srinivasan:1998ty,Shankaranarayanan:2000gb}. In this point, it would be interesting to calculate the thermodynamic quantities of black holes in tunneling formalism with GUP. Recently, such an investigation has been highly regarded in the literature~\cite{Tawfik:2015kga,Gangopadhyay:2014wja,Anacleto:2015kca,Majumder:afa,Anacleto:2015awa,Anacleto:2015mma,Sakalli:2016mnk,Faizal:2014tea}.

All above information give us motivation to use tunneling formalism with GUP and determine the thermodynamic properties of the charged non-rotating BTZ black hole. In the section \ref{sec:2}, the quantum tunneling phenomena by WKB method in the near horizon of the charged non-rotating BTZ black hole is reviewed. In the section \ref{sec:3}, the quantum corrections to thermodynamic quantities of the charged non-rotating BTZ black hole, due to the effects of GUP in tunneling formalism, are investigated\footnote{The authors would like to mention that such an investigation for the BTZ black hole was presented in ~\cite{Modak:2008tg,Akbar:2010nq} by an alternative approach.}. In the section \ref{sec:4}, some conclusions and discussions are presented.

\section{Charged non-rotating BTZ black hole}
\label{sec:2}
The Hilbert action in the 2+1 dimensional Einstein-Maxwell gravity with a negative cosmological constant is given by
\begin{equation}\label{2.1}
  S = \frac{1}{\pi} \int d^{3}x \sqrt{-g} ( \frac{R - 2 \Lambda}{16 G_{3}} - \frac{1}{4} F^{2}) ,
\end{equation}
where $G_{3}$, $\Lambda = - \frac{1}{\ell^{2}}$ and $ F = dA $ are the Newton's gravitational constant in 2+1 dimensions, the cosmological constant and the Maxwell strength tensor, respectively. The charged non-rotating BTZ black hole is a static solution to the above action. In units $G_{3} = \frac{1}{8}$, the metric is given by~\cite{Banados:1992wn,Martinez:1999qi,Clement:1995zt}
\begin{equation}\label{2.2}
  ds^{2} = - ( \frac{r^{2}}{\ell^{2}} - M - \frac{Q^{2}}{2} \ln(\frac{r^{2}}{\ell^{2}})) dt^{2} + \frac{1}{\frac{r^{2}}{\ell^{2}} - M - \frac{Q^{2}}{2} \ln(\frac{r^{2}}{\ell^{2}})} dr^{2} + r^{2} d\phi^{2} ,
\end{equation}
where $M$, $Q$ and $\ell$ are ADM mass, charge of the black hole and $AdS$ radius, respectively. The corresponding non-vanishing component of the electromagnetic field is given by
\begin{equation}\label{2.3}
 A_{t} = -Q \ln(\frac{r}{\ell}) .\\
\end{equation}
The horizons are given by
\begin{equation}\label{2.4}
 \frac{r_{\pm}^{2}}{\ell^{2}} - M - \frac{Q^{2}}{2} \ln(\frac{r_{\pm}^{2}}{\ell^{2}})=0 ,
\end{equation}
where $r_{+}$ and $r_{-}$ present outer (event) and inner (Cauchy) horizon, respectively. The Hawking temperature and Entropy are ,respectively, given by
\begin{equation}\label{2.5}
 T_{H}=T_{+}=\frac{\kappa_{+}}{2\pi}=\frac{1}{2\pi}(\frac{r_{+}}{\ell^{2}}-\frac{Q^{2}}{2 r_{+}}) ,
\end{equation}
\begin{equation}\label{2.6}
 S_{+}=4 \pi r_{+} .
\end{equation}
In equation \eqref{2.5}, $\kappa_{+}$ is the surface gravity at the outer horizon.

To study the black hole thermodynamics by Hamilton-Jacobi method, one needs to find the near horizon limit of black hole. So taking the near (outer) horizon limit, the near (outer) horizon metric is obtained as following
\begin{equation}\label{2.7}
 ds^{2}_{NH}=- \left( (\frac{2r_{+}}{\ell^{2}}-\frac{Q^{2}}{r_{+}}) (r-r_{+}) \right) dt^{2} + \left( (\frac{2r_{+}}{\ell^{2}}-\frac{Q^{2}}{r_{+}}) (r-r_{+}) \right)^{-1} dr^{2} + r_{+}^{2} d\phi^{2} .
\end{equation}
Now, one may consider massive scalar particles propagating in the background of \eqref{2.7},
\begin{equation}\label{2.8}
  \left(\frac{1}{\sqrt{-g}}\partial_{\mu}(\sqrt{-g} g^{\mu\nu} \partial_{\nu})+m^{2}\right)\Phi = 0 ,
\end{equation}
where $m$ is the mass of scalar particles. Taking $\Phi=\exp( i I(t,r,\phi) )$, the above massive Klein-Gordon equation in the background of \eqref{2.7} is simplified to
\begin{equation}\label{2.9}
  -\frac{1}{(\frac{2r_{+}}{\ell^{2}}-\frac{Q^{2}}{r_{+}}) (r-r_{+})} (\partial_{t}I)^{2}+((\frac{2r_{+}}{\ell^{2}}-\frac{Q^{2}}{r_{+}}) (r-r_{+})) (\partial_{r}I)^{2}+\frac{1}{r_{+}^{2}}(\partial_{\phi}I)^{2}+m^{2}=0 .
\end{equation}
Taking the following ansatz
\begin{equation}\label{2.10}
  I=-Et+W(r)+J_{m}\phi
\end{equation}
where $J_{m}$ is the constant angular momentum of particles, one can easily obtain
 \begin{equation}\label{2.11}
  -\frac{1}{(\frac{2r_{+}}{\ell^{2}}-\frac{Q^{2}}{r_{+}}) (r-r_{+})} (-E)^{2}+((\frac{2r_{+}}{\ell^{2}}-\frac{Q^{2}}{r_{+}}) (r-r_{+})) (\frac{dW(r)}{dr})^{2}+\frac{1}{r_{+}^{2}}(J_{m})^{2}+m^{2}=0 .
\end{equation}
Solving above equation leads to
\begin{equation}\label{2.12}
  W(r)=\int \frac{\sqrt{E^{2}-((\frac{2r_{+}}{\ell^{2}}-\frac{Q^{2}}{r_{+}}) (r-r_{+}))(\frac{(J_{m})^{2}}{r_{+}^{2}}+m^{2})}}{(\frac{2r_{+}}{\ell^{2}}-\frac{Q^{2}}{r_{+}}) (r-r_{+})}dr=\frac{\pi i E}{\kappa_{+}} .
\end{equation}
The particles tunneling rate is given by $\Gamma \sim e^{-2Im(I)}$, so in the present case it's given by
 \begin{equation}\label{2.13}
   \Gamma \sim e^{-\frac{2 \pi E}{\kappa_{+}}} .
 \end{equation}
Now the Hawking temperature of the charged non-rotating BTZ black hole can be got comparing the equation \eqref{2.13} with Boltzmann distribution $e^{-\frac{E}{T}}$, so the Hawking temperature would be
 \begin{equation}\label{2.14}
   T_{H}=\frac{\kappa_{+}}{2\pi}=\frac{1}{2\pi}(\frac{r_{+}}{\ell^{2}}-\frac{Q^{2}}{2 r_{+}}) .
 \end{equation}

\section{Quantum corrections to BTZ black hole thermodynamics}
\label{sec:3}

To investigate quantum corrections to the BTZ black hole thermodynamics, one may consider the GUP effects in the tunneling formalism. To this end, one can consider the GUP relation \eqref{1.2} and write it as follows
\begin{equation}\label{3.1}
  \Delta p \geq \frac{\hbar (\Delta x+\lambda l_{p})}{2 \lambda^{2} l_{p}^{2}} \left(1-\sqrt{1-\frac{4 \lambda^{2} l_{p}^{2}}{(\Delta x+\lambda l_{p})^{2}}}\right) .
\end{equation}
Taking Taylor expansion around $\frac{\lambda l_{p}}{\Delta x}$, the inequation \eqref{3.1} would become as follows
\begin{equation}\label{3.2}
  \Delta p \geq \frac{1}{\Delta x} \left( 1 - \frac{\lambda l_{p}}{\Delta x} + \frac{2 \lambda^{2} l_{p}^{2}}{\Delta x^{2}}+\cdots \right)
\end{equation}
 where $\hbar=1$ has been chosen. Now using the saturated form of the uncertainty principle $E \Delta x \geq 1$, which follows from the saturated form of the Heisenberg uncertainty principle $\Delta x \Delta p \geq 1$~\cite{Gangopadhyay:2014wja}, in the inequation \eqref{3.2}, one can get
 \begin{equation}\label{3.3}
  E_{GUP} \geq E \left( 1 - \frac{\lambda l_{p}}{\Delta x} + \frac{2 \lambda^{2} l_{p}^{2}}{\Delta x^{2}} \right)
\end{equation}
where written up to the second order in $l_{p}$. Here $E$ is the energy of the tunneling particles and $E_{GUP}$ is the corrected energy of them. Like section \ref{sec:2}, the tunneling rate of particles with energy $E_{GUP}$ reads
\begin{equation}\label{3.4}
   \Gamma \sim e^{-\frac{2 \pi E_{GUP}}{\kappa_{+}}} .
 \end{equation}
As discussed in section \ref{sec:2}, comparing the equation \eqref{3.4} with the Boltzmann distribution $e^{-\frac{E}{T}}$, the quantum corrected Hawking temperature can be obtained
\begin{equation}\label{3.5}
   T_{GUP}=\frac{\kappa_{+}}{2 \pi} \left( 1 - \frac{\lambda l_{p}}{\Delta x} + \frac{2 \lambda^{2} l_{p}^{2}}{\Delta x^{2}} \right)^{-1} .
 \end{equation}
Now it may be chosen $\Delta x=2r_{+}$, by this fact that the uncertainty in the position of the particle near the black
hole horizon is of order of the horizon radius~\cite{Adler:2001vs}. Therefore, the quantum corrected Hawking temperature can be rewritten as
\begin{equation}\label{3.6}
   T_{GUP}=T_{H} \left( 1 - \frac{\lambda l_{p}}{2 r_{+}} + \frac{ \lambda^{2} l_{p}^{2}}{2 r_{+}^{2}} \right)^{-1},
 \end{equation}
or
\begin{equation}\label{3.7}
   T_{GUP} \simeq T_{H} \left( 1 + \frac{\lambda l_{p}}{2 r_{+}} - \frac{ \lambda^{2} l_{p}^{2}}{2 r_{+}^{2}} \right).
 \end{equation}

To determine the quantum corrected black hole entropy, one can use the first law of black hole thermodynamics
given by
\begin{equation}\label{3.8}
   S_{GUP}=\int \frac{dM}{T_{GUP}}.
 \end{equation}
Putting \eqref{3.6} into \eqref{3.8} and carrying out the integral leads to
\begin{equation}\label{3.9}
   S_{GUP} = 4 \pi r_{+} - 2 \pi \lambda l_{p} \ln(r_{+}) - \frac{2 \pi \lambda^{2} l_{p}^{2}}{r_{+}} + constant .
 \end{equation}
The quantum corrected black hole entropy can be rewritten as
\begin{equation}\label{3.10}
   S_{GUP} = S - 2 \pi \lambda l_{p} \ln(\frac{S}{4 \pi}) - \frac{8 \pi^{2} \lambda^{2} l_{p}^{2}}{S} + constant ,
 \end{equation}
 or equivalently
 \begin{equation}\label{3.11}
   S_{GUP} = S - 2 \pi \lambda l_{p} \ln S - \frac{8 \pi^{2} \lambda^{2} l_{p}^{2}}{S} + \overline{constant} .
 \end{equation}
As it is seen, the quantum corrected entropy of the BTZ black hole in tunneling formalism with GUP, \eqref{3.11}, is of the form of the predicted entropy by string theory and loop quantum gravity, \eqref{1.1}. The modified entropy \eqref{3.11}, which has been obtained by the most general form of GUP \eqref{1.2}, is the same with that of~\cite{Medved:2004yu} which was obtained by GUP including just a quadratic term in the momenta; but two points should be noted. First, it is expected that all of GUPs lead to the same result, hence it depends on the objectives of the study and the problem. For example, in the present study, it is intended to work with the general form of GUP which is compatible with all well-known theories of quantum gravity, such as string theory, loop quantum gravity and DSR. Second, the expansion \eqref{3.2}, which governs the result, leads to the same terms in both GUP approaches. So, it is acceptable that these two different versions of GUP present the same result.\\

The heat specific capacity at constant charge, $C_{Q}$, is given by
  \begin{equation}\label{3.12}
   C_{Q} = T_{H} \left(\frac{\partial S}{\partial T_{H}}\right)_{Q} ,
 \end{equation}
 calculating above equation leads to
 \begin{equation}\label{3.13}
   C_{Q} = \frac{8 \pi^{2} \ell^{2} T_{H}}{1+\frac{Q^{2} \ell^{2}}{2 r_{+}^{2}}} .
 \end{equation}
 Note that putting $Q=0$, the equation \eqref{3.13} gives the specific heat capacity of non-charged non-rotating BTZ black hole. In analogy with classical case, the quantum corrected specific heat capacity at constant charge, $\left(C_{GUP}\right)_{Q}$, is given by
 \begin{equation}\label{3.14}
  \left(C_{GUP}\right)_{Q} = T_{GUP} \left(\frac{\partial S_{GUP}}{\partial T_{GUP}}\right)_{Q} .
 \end{equation}
 Finally, doing some calculation and keeping correction terms up to second order in Planck length, the quantum corrections to the specific heat capacity at constant charge is obtained
 \begin{equation}\label{3.15}
   \left(C_{GUP}\right)_{Q} \simeq \frac{ 1 - \left( \frac{ \lambda l_{p}}{2 r_{+}} - \frac{ \lambda^{2} l_{p}^{2}}{2 r_{+}^{2}} \right)}{1- \left( \frac{\lambda l_{p}}{2 r_{+}} - \frac{3 \lambda^{2} l_{p}^{2}}{4 r_{+}^{2}} \right) \varepsilon} C_{Q} ,
 \end{equation}
 where $\varepsilon$ is given by
\begin{equation}\label{3.16}
  \varepsilon = \frac{1-\frac{Q^{2} \ell^{2}}{2 r_{+}^{2}}}{1+\frac{Q^{2} \ell^{2}}{2 r_{+}^{2}}} .
\end{equation}
As expected, the quantum corrections are so small while $\frac{\lambda l_{p}}{\Delta x} \ll 1$.

\section{Conclusions}
\label{sec:4}

Using quantum tunneling formalism in presence of generalized uncertainty principle, thermodynamics quantities of the charged non-rotating BTZ has been investigated. It has been shown that the black hole entropy is of the form of the predicted entropy by string theory and loop quantum gravity. It should be mentioned, since it is assumed that $\frac{\lambda l_{p}}{\Delta x} \ll 1$, thermodynamics of the black hole can not be studied for very small $r_{+}$. In that case, when the event horizon radius approaches zero, the quantum corrections become significant, however, there would exist a point such that the approximation \eqref{3.2} would not be valid anymore. As final note, It would be of interest to study the charged non-rotating BTZ black hole in the noncommutative spacetime via tunneling formalism with GUP. Such an investigation may lead to more correction terms to the thermodynamics of the charged non-rotating BTZ black hole.

\section{Acknowledgment}
The authors would like to thank the anonymous referee for his/her valuable comments and suggestions to improve the quality of the paper.


\bibliographystyle{JHEP.bst}
\bibliography{paper2016930}

\providecommand{\href}[2]{#2}\begingroup\raggedright\begin{thebibliography}{10}

\bibitem{Bekenstein:1972tm}
J.~D. Bekenstein, {\it {Black holes and the second law}},  {\em Lett. Nuovo
  Cim.} {\bf 4} (1972) 737--740.

\bibitem{Bekenstein:1973ur}
J.~D. Bekenstein, {\it {Black holes and entropy}},  {\em Phys. Rev.} {\bf D7}
  (1973) 2333--2346.

\bibitem{Bekenstein:1974ax}
J.~D. Bekenstein, {\it {Generalized second law of thermodynamics in black hole
  physics}},  {\em Phys. Rev.} {\bf D9} (1974) 3292--3300.

\bibitem{Bardeen:1973gs}
J.~M. Bardeen, B.~Carter, and S.~W. Hawking, {\it {The Four laws of black hole
  mechanics}},  {\em Commun. Math. Phys.} {\bf 31} (1973) 161--170.

\bibitem{Hawking:1974rv}
S.~W. Hawking, {\it {Black hole explosions}},  {\em Nature} {\bf 248} (1974)
  30--31.

\bibitem{Hawking:1974sw}
S.~W. Hawking, {\it {Particle Creation by Black Holes}},  {\em Commun. Math.
  Phys.} {\bf 43} (1975) 199--220.

\bibitem{Carlip:2000nv}
S.~Carlip, {\it {Logarithmic corrections to black hole entropy from the Cardy
  formula}},  {\em Class. Quant. Grav.} {\bf 17} (2000) 4175--4186,
  [\href{http://arxiv.org/abs/gr-qc/0005017}{{\tt gr-qc/0005017}}].

\bibitem{Kaul:2000kf}
R.~K. Kaul and P.~Majumdar, {\it {Logarithmic correction to the
  Bekenstein-Hawking entropy}},  {\em Phys. Rev. Lett.} {\bf 84} (2000)
  5255--5257, [\href{http://arxiv.org/abs/gr-qc/0002040}{{\tt gr-qc/0002040}}].

\bibitem{Medved:2004yu}
A.~J.~M. Medved and E.~C. Vagenas, {\it {When conceptual worlds collide: The
  GUP and the BH entropy}},  {\em Phys. Rev.} {\bf D70} (2004) 124021,
  [\href{http://arxiv.org/abs/hep-th/0411022}{{\tt hep-th/0411022}}].

\bibitem{Myung:2003ra}
Y.~S. Myung, {\it {Logarithmic corrections to three-dimensional black holes and
  de Sitter spaces}},  {\em Phys. Lett.} {\bf B579} (2004) 205--210,
  [\href{http://arxiv.org/abs/hep-th/0310176}{{\tt hep-th/0310176}}].

\bibitem{Meissner:2004ju}
K.~A. Meissner, {\it {Black hole entropy in loop quantum gravity}},  {\em
  Class. Quant. Grav.} {\bf 21} (2004) 5245--5252,
  [\href{http://arxiv.org/abs/gr-qc/0407052}{{\tt gr-qc/0407052}}].

\bibitem{Faizal:2014rwa}
M.~Faizal and B.~Majumder, {\it {Incorporation of Generalized Uncertainty
  Principle into Lifshitz Field Theories}},  {\em Annals Phys.} {\bf 357}
  (2015) 49--58, [\href{http://arxiv.org/abs/1408.3795}{{\tt
  arXiv:1408.3795}}].

\bibitem{Pramanik:2014mma}
S.~Pramanik, M.~Faizal, M.~Moussa, and A.~F. Ali, {\it {Path integral
  quantization corresponding to the deformed Heisenberg algebra}},  {\em Annals
  Phys.} {\bf 362} (2015) 24--35, [\href{http://arxiv.org/abs/1411.4979}{{\tt
  arXiv:1411.4979}}].

\bibitem{Ali:2015ola}
A.~F. Ali, M.~Faizal, and M.~M. Khalil, {\it {Short Distance Physics of the
  Inflationary de Sitter Universe}},  {\em JCAP} {\bf 1509} (2015), no.~09 025,
  [\href{http://arxiv.org/abs/1505.06963}{{\tt arXiv:1505.06963}}].

\bibitem{Garattini:2015aca}
R.~Garattini and M.~Faizal, {\it {Cosmological constant from a deformation of
  the Wheeler–DeWitt equation}},  {\em Nucl. Phys.} {\bf B905} (2016)
  313--326, [\href{http://arxiv.org/abs/1510.04423}{{\tt arXiv:1510.04423}}].

\bibitem{Faizal:2016zlo}
M.~Faizal, {\it {Supersymmetry breaking as a new source for the generalized
  uncertainty principle}},  {\em Phys. Lett.} {\bf B757} (2016) 244--246,
  [\href{http://arxiv.org/abs/1605.00925}{{\tt arXiv:1605.00925}}].

\bibitem{Faizal:2014pia}
M.~Faizal, {\it {Consequences of Deformation of the Heisenberg Algebra}},  {\em
  Int. J. Geom. Meth. Mod. Phys.} {\bf 12} (2014), no.~02 1550022,
  [\href{http://arxiv.org/abs/1404.5024}{{\tt arXiv:1404.5024}}].

\bibitem{Faizal:2014dua}
M.~Faizal, A.~F. Ali, and A.~Nassar, {\it {AdS/CFT Correspondence Beyond its
  Supergravity Approximation}},  {\em Int. J. Mod. Phys.} {\bf A30} (2015),
  no.~30 1550183, [\href{http://arxiv.org/abs/1405.4519}{{\tt
  arXiv:1405.4519}}].

\bibitem{Faizal:2014rha}
M.~Faizal, {\it {Deformation of the Wheeler-DeWitt Equation}},  {\em Int. J.
  Mod. Phys.} {\bf A29} (2014), no.~20 1450106,
  [\href{http://arxiv.org/abs/1406.0273}{{\tt arXiv:1406.0273}}].

\bibitem{Faizal:2014mfa}
M.~Faizal and S.~I. Kruglov, {\it {Deformation of the Dirac Equation}},  {\em
  Int. J. Mod. Phys.} {\bf D25} (2015), no.~01 1650013,
  [\href{http://arxiv.org/abs/1406.2653}{{\tt arXiv:1406.2653}}].

\bibitem{Faizal:2014mba}
M.~Faizal, M.~M. Khalil, and S.~Das, {\it {Time Crystals from Minimum Time
  Uncertainty}},  {\em Eur. Phys. J.} {\bf C76} (2016), no.~1 30,
  [\href{http://arxiv.org/abs/1501.03111}{{\tt arXiv:1501.03111}}].

\bibitem{Ali:2009zq}
A.~F. Ali, S.~Das, and E.~C. Vagenas, {\it {Discreteness of Space from the
  Generalized Uncertainty Principle}},  {\em Phys. Lett.} {\bf B678} (2009)
  497--499, [\href{http://arxiv.org/abs/0906.5396}{{\tt arXiv:0906.5396}}].

\bibitem{Ali:2011fa}
A.~F. Ali, S.~Das, and E.~C. Vagenas, {\it {A proposal for testing Quantum
  Gravity in the lab}},  {\em Phys. Rev.} {\bf D84} (2011) 044013,
  [\href{http://arxiv.org/abs/1107.3164}{{\tt arXiv:1107.3164}}].

\bibitem{Parikh:1999mf}
M.~K. Parikh and F.~Wilczek, {\it {Hawking radiation as tunneling}},  {\em
  Phys. Rev. Lett.} {\bf 85} (2000) 5042--5045,
  [\href{http://arxiv.org/abs/hep-th/9907001}{{\tt hep-th/9907001}}].

\bibitem{Srinivasan:1998ty}
K.~Srinivasan and T.~Padmanabhan, {\it {Particle production and complex path
  analysis}},  {\em Phys. Rev.} {\bf D60} (1999) 024007,
  [\href{http://arxiv.org/abs/gr-qc/9812028}{{\tt gr-qc/9812028}}].

\bibitem{Shankaranarayanan:2000gb}
S.~Shankaranarayanan, K.~Srinivasan, and T.~Padmanabhan, {\it {Method of
  complex paths and general covariance of Hawking radiation}},  {\em Mod. Phys.
  Lett.} {\bf A16} (2001) 571--578,
  [\href{http://arxiv.org/abs/gr-qc/0007022}{{\tt gr-qc/0007022}}].

\bibitem{Tawfik:2015kga}
A.~N. Tawfik and E.~A. El~Dahab, {\it {Corrections to entropy and
  thermodynamics of charged black hole using generalized uncertainty
  principle}},  {\em Int. J. Mod. Phys.} {\bf A30} (2015), no.~09 1550030,
  [\href{http://arxiv.org/abs/1501.01286}{{\tt arXiv:1501.01286}}].

\bibitem{Gangopadhyay:2014wja}
S.~Gangopadhyay, {\it {Minimal length effects in black hole thermodynamics from
  tunneling formalism}},  {\em Int. J. Theor. Phys.} {\bf 55} (2016), no.~1
  617--624, [\href{http://arxiv.org/abs/1405.4229}{{\tt arXiv:1405.4229}}].

\bibitem{Anacleto:2015kca}
M.~A. Anacleto, F.~A. Brito, A.~G. Cavalcanti, E.~Passos, and J.~Spinelly, {\it
  {Quantum correction to the entropy of noncommutative BTZ black hole}},
  \href{http://arxiv.org/abs/1510.08444}{{\tt arXiv:1510.08444}}.

\bibitem{Majumder:afa}
B.~Majumder, {\it {Black Hole Entropy with minimal length in Tunneling
  formalism}},  {\em Gen. Rel. Grav.} {\bf 45} (2013) 2403--2414,
  [\href{http://arxiv.org/abs/1212.6591}{{\tt arXiv:1212.6591}}].

\bibitem{Anacleto:2015awa}
M.~A. Anacleto, F.~A. Brito, G.~C. Luna, E.~Passos, and J.~Spinelly, {\it
  {Quantum-corrected finite entropy of noncommutative acoustic black holes}},
  {\em Annals Phys.} {\bf 362} (2015) 436--448,
  [\href{http://arxiv.org/abs/1502.00179}{{\tt arXiv:1502.00179}}].

\bibitem{Anacleto:2015mma}
M.~A. Anacleto, F.~A. Brito, and E.~Passos, {\it {Quantum-corrected self-dual
  black hole entropy in tunneling formalism with GUP}},  {\em Phys. Lett.} {\bf
  B749} (2015) 181--186, [\href{http://arxiv.org/abs/1504.06295}{{\tt
  arXiv:1504.06295}}].

\bibitem{Sakalli:2016mnk}
I.~Sakalli, A.~Ovgun, and K.~Jusufi, {\it {GUP Assisted Hawking Radiation of
  Rotating Acoustic Black Holes}},  \href{http://arxiv.org/abs/1602.04304}{{\tt
  arXiv:1602.04304}}.

\bibitem{Faizal:2014tea}
M.~Faizal and M.~M. Khalil, {\it {GUP-Corrected Thermodynamics for all Black
  Objects and the Existence of Remnants}},  {\em Int. J. Mod. Phys.} {\bf A30}
  (2015), no.~22 1550144, [\href{http://arxiv.org/abs/1411.4042}{{\tt
  arXiv:1411.4042}}].

\bibitem{Modak:2008tg}
S.~K. Modak, {\it {Corrected entropy of BTZ black hole in tunneling approach}},
   {\em Phys. Lett.} {\bf B671} (2009) 167--173,
  [\href{http://arxiv.org/abs/0807.0959}{{\tt arXiv:0807.0959}}].

\bibitem{Akbar:2010nq}
M.~Akbar and K.~Saifullah, {\it {Quantum corrections to the entropy of charged
  rotating black holes}},  {\em Eur. Phys. J.} {\bf C67} (2010) 205--211,
  [\href{http://arxiv.org/abs/1002.3581}{{\tt arXiv:1002.3581}}].

\bibitem{Banados:1992wn}
M.~Banados, C.~Teitelboim, and J.~Zanelli, {\it {The Black hole in
  three-dimensional space-time}},  {\em Phys. Rev. Lett.} {\bf 69} (1992)
  1849--1851, [\href{http://arxiv.org/abs/hep-th/9204099}{{\tt
  hep-th/9204099}}].

\bibitem{Martinez:1999qi}
C.~Martinez, C.~Teitelboim, and J.~Zanelli, {\it {Charged rotating black hole
  in three space-time dimensions}},  {\em Phys. Rev.} {\bf D61} (2000) 104013,
  [\href{http://arxiv.org/abs/hep-th/9912259}{{\tt hep-th/9912259}}].

\bibitem{Clement:1995zt}
G.~Clement, {\it {Spinning charged BTZ black holes and selfdual particle - like
  solutions}},  {\em Phys. Lett.} {\bf B367} (1996) 70--74,
  [\href{http://arxiv.org/abs/gr-qc/9510025}{{\tt gr-qc/9510025}}].

\bibitem{Adler:2001vs}
R.~J. Adler, P.~Chen, and D.~I. Santiago, {\it {The Generalized uncertainty
  principle and black hole remnants}},  {\em Gen. Rel. Grav.} {\bf 33} (2001)
  2101--2108, [\href{http://arxiv.org/abs/gr-qc/0106080}{{\tt gr-qc/0106080}}].

\end{thebibliography}\endgroup


\end{document}